\documentstyle[12pt]{article}
\begin{document}
\font\bss=cmr12 scaled\magstep 0
\title{Parasupersymmetric structure of the Boussinesq-type systems }
\author{A.V. Yurov\\
\small  Theoretical Physics Department,\\
\small  Kaliningrad State
University, \\
\small 236041, Al.Nevsky St., 14, Kaliningrad, Russia \\
\small   email yurov@freemail.ru}
\date {}
\renewcommand{\abstractname}{\small Abstract}
\maketitle
\begin{abstract}
We study Darboux transformations for a Boussinesq-type equations. The parasupersymmetric 
structure of link between Boussinesq and modified Boussinesq systems is revealed.
\end{abstract}
\thispagestyle{empty}
\medskip
\section{Introduction.}
In [1] the supersymmetric structure of the KdV and modified KdV (mKdV) systems 
(including lower KdV equations) is revealed. Well known Miura transformations 
between KdV and mKdV equations is nothing but the manifestation of this structure.   
In [2] the Miura-type transformations for the Boussinesq (Bq) and modified Boussinesq 
(mBq) equations is obtained. Lax pairs for these equations are some third-order linear 
differential expressions and its can't be elements of supersymmetry (SUSY) algebra. 
This because SUSY algebra can be realized via even-order linear operators [3]. 

Some times ago interest has appeared into extensions of SUSY quantum mechanics to systems 
with three-fold degeneracy of the energy spectrum [4-6]. The related transformations obey 
the {\it parasuperalgebra}. By contrast with SUSY which bind one bosonic and one 
fermionic levels with the same energy, the parasupersymmetry (PSUSY) do the same  
for the one bosonic and two parafermionic levels [7]. 

One of our main goals is to show that the algebraic structure of Bq and mBq equations 
is the PSUSY and that Miura-type transformations between these systems ([2]) can be  
obtained from this structure. 
\section{Supersymmetry and parasupersymmetry.}
The KdV equation can be obtained from the Lax pair in the form 
$$
\frac{dL_1}{dt}=\left[A_1,L_1\right],
\eqno(1)
$$
where $L_1\equiv \partial^2+u_1(x,t)=q^+q$, $\partial=d/dx$, $q=\partial+g$, $q^+=\partial-g$.

Darboux transformation (DT) for the Schr\"odinger equation yet [8],
$$
L_1\to L_2=\partial^2+u_2=qq^+,
\eqno(2)
$$
with  $u_2=u_1+2g_x$. 

Supersymmetry Hamiltonian $H$ and supergenerators $Q$, $Q^+$ are [9],
$$
H=\left(\begin{array}{cc}
L_1&0\\
0&L_2
\end{array}\right),
\qquad 
Q=\left(\begin{array}{cc}
0&0\\
q&0
\end{array}\right),
\qquad
Q^+=\left(\begin{array}{cc}
0&q^+\\
0&0
\end{array}\right),
$$
so we have superalgebra
$$
\{Q,Q^+\}=H,\qquad [Q,H]=[Q^+,H]=0.
\eqno(3)
$$
Note the supergenerators are nilpotent of order two,
$$
Q^2=\left(Q^+\right)^2=0.
$$
To construct PSUSY we use two-times DT
$$
L_1=q^+q\to L_2=qq^+={\tilde q}^+{\tilde q}+\mu\to L_3={\tilde q}{\tilde q}^++\mu,
$$
where ${\tilde q}=\partial +{\tilde g}$,  ${\tilde q}^+=\partial -{\tilde g}$. Then the 
parasuperhamiltonian and parasupergenerators are 
$$
H=\left(\begin{array}{ccc}
L_1&0&0\\
0&L_2&0\\
0&0&L_3-\mu
\end{array}\right),
\qquad 
Q=\left(\begin{array}{ccc}
0&0&0\\
q&0&0\\
0&{\tilde q}&0
\end{array}\right),
\qquad
Q^+=\left(\begin{array}{ccc}
0&q^+&0\\
0&0&{\tilde q}^+\\
0&0&0
\end{array}\right).
\eqno(4)
$$
Parasupergenerators are nilpotent of order three,
$$
Q^3=\left(Q^+\right)^3=0,
$$
and they satisfy parasuperalgebra [6],
$$
Q^+QQ^+=Q^+H,\qquad (Q^+)^2Q+Q(Q^+)^2=Q^+H,\qquad
[H,Q]=[H,Q^+]=0,
\eqno(5)
$$
if $\mu=0$. 

{\bf Remark.}  The fact that $\mu=0$ show that only special kind of DT - {\it binary DT} - 
is working  to construct parasuperalgebra. In fact, $g=-\psi_x/\psi$, where $L_1\psi=0$ 
and ${\tilde g}=-{\tilde \psi_x}/{\tilde \psi}$, where $L_2{\tilde\psi}=\mu{\tilde\psi}$. 
Therefore for the case  $\mu=0$ there are only two variants for the $\tilde\psi$. 
The first is $\tilde\psi=1/\psi$. In this case $u_3=L_3-\partial^2=u_1$ so $L_1=L_3$. It is 
trivial and uninteresting case. 

The second variant for the $\tilde\psi$ is 
$$
\tilde\psi=\frac{1}{\psi}\int dx\psi^2,
$$
so 
$$
u_3=u_1+2\partial^2\log\int dx\psi^2.
\eqno(6)
$$
(6) is so called binary DT (it is  the DT to square). It is fundamental relationship in the 
positon theory [10]. In particular, one can show that one-positon (or one-negaton) solution 
of the KdV equation can be obtained via the formula (6) if $u_1=0$. Thus, positon potentials 
are connected with the PSUSY whereas solitons are the same for the SUSY.   
\section{PSUSY structure of integrable systems.}
Andreev and Burova showed that connection between KdV and mKdV equations has SUSY 
structure [1]. To show this one need to construct supercharge  which is nothing but 
square root from the SUSY Hamiltonian~\footnote{There are two such operators: 
$\theta$ and $\theta'=i\sigma_3\theta$, where $\sigma_3$ is Pauli  matrix.}:
$$
\theta=\sqrt{H}.
$$
$\theta$ is $2\times 2$ matrix operator and {\it  it is the $L$-operator of the mKdV hierarchy.} 
This is the crucial point of the work [1]. 

Now let consider Bq system 
$$
a_{1t}=(2b_1-a_{1x})_x,\qquad b_{1t}=\left(b_{1x}-\frac{2}{3}a_{1xx}-
\frac{1}{3}a_1^2\right)_x,
\eqno(7)
$$
and a certain "modified" version of it given by  ([2])
$$
\begin{array}{l}
f_{1t}=f_{1xx}-2f_1f_{1x}-\frac{2}{3}\left(2f_1+f_2\right)_{xx}-
\frac{2}{3}\left(f_1f_2-\left(f_1+f_2\right)^2\right)_x,\\
f_{2t}=f_{2xx}-2f_2f_{2x}-\frac{2}{3}\left(f_2-f_1\right)_{xx}-
\frac{2}{3}\left(f_1f_2-\left(f_1+f_2\right)^2\right)_x.
\end{array}
\eqno(8)
$$
One of the goal of the work [2] was to relate solutions $a_1$, $b_1$ of (7) and 
$f_1$, $f_2$ of (8) by Miura-type transformation. Authors did it but what is the 
algebraic structure which allow one to obtain this Miura-type transformation? Our aim 
here is to show that it is possible because (7) and (8) are connected via PSUSY. 
To show this we are starting with Lax representation (1) for the (7) where 
$$
L_1=\partial^3+a_1\partial+b_1,\qquad A_1=\partial^2+\frac{2}{3}a_1.
$$
It is well known that ([2]),
$$
L_1=\left(\partial+f_3\right) \left(\partial+f_2\right) \left(\partial+f_1\right)=q_3q_2q_1,
$$
where 
$$
\begin{array}{l}
f_1+f_2+f_3=0,\\
a_1=\left(f_2+2f_1\right)_x-f_1^2-f_2^2-f_1f_2,\\
b_1=f_{1xx}+f_1\left(f_2-f_1\right)_x-f_1f_2\left(f_1+f_2\right).
\end{array}
\eqno(9)
$$
By the analogy with (2) we get two DT
$$
L_1\to L_2\to L_3,
$$
or 
$$
q_3q_2q_1\to q_1q_3q_2\to q_2q_1q_3,
$$
where 
$$
L_2=\partial^3+a_2\partial+b_2, \qquad L_3=\partial^3+a_3\partial+b_3,
$$
with 
$$
a_2=a_1-3f_{1x},\qquad a_3=a_2-3f_{2x}.
\eqno(10)
$$
We don't need $b_2$ and $b_3$. 

As for the usual SUSY one can construct the first parasuperhamiltonian,
$$
H_{_I}=\left(\begin{array}{ccc}
L_1&0&0\\
0&L_2&0\\
0&0&L_3
\end{array}\right).
\eqno(11)
$$
To contrast with (4), $L_i$ ($i=1,2,3$) are third-order linear differential operators. 

To construct parasupercharge $M$ we must calculate the cube root from the (11): 
$M=H^{1/3}$. It  easy to verify that 
$$
M=\left(\begin{array}{ccc}
0&0&q_3\\
q_1&0&0\\
0&q_2&0
\end{array}\right).
\eqno(12)
$$
The rest roots can be obtained by the multiplication of $M$ to the matrix 
$$
\left(\begin{array}{ccc}
\lambda_1&0&0\\
0&\lambda_2&0\\
0&0&(\lambda_1\lambda_2)^{-1}
\end{array}\right),
$$
where $\lambda_{1,2}$  are arbitrary (nonvanishing) complex numbers. 

Operator (12) is contained in [2]. Namely, the Lax equation for the (8) is 
$$
\frac{dM}{dt}=[H_{_{II}},M],
$$
thus parasupercharge $M$ is the $L$-operator for the (8). Therefore it is clear 
why we can call (8) as modified Bq system. The $A$-operator $H_{_{II}}$ has the form 
$$
H_{_{II}}=\left(\begin{array}{ccc}
A_1&0&0\\
0&A_2&0\\
0&0&A_3
\end{array}\right).
\eqno(13)
$$
where $A_i=\partial^2+\frac{2}{3}a_i$ (see (10)). So (13) look as  the 
parasuperhamiltonian (4) exactly. To show that (13)=(4) (with $\mu=0$) we need to 
find two functions $g$ and $\tilde g$ such that 
$$
A_1=(\partial-g)(\partial+g),\qquad 
A_2=(\partial+g)(\partial-g)= (\partial-{\tilde g})(\partial+{\tilde g}),\qquad 
A_3=(\partial+{\tilde g})(\partial-{\tilde g}).
$$
Using (9) one get 
$$
g=f_1+c_1,\qquad {\tilde g}=f_2+c_2,
$$
with some constants $c_1$ and $c_2$. In this case functions $f_1$ and $f_2$ are not arbitrary.  
After calculation we get the nonlinear equation for the $f_1$,
$$
\begin{array}{l}
2(2c_2-f_1)\left(f_{1x}+2(c_2-c_1)f_1\right)_x+f_{1x}^2+\\
\left(\left(f_1+2c_1-c_2\right)^2-
3\left(c_1^2+c_2^2\right)\right)\left(3(f_1-c_2)^2-(c_1+c_2)^2-2c_1c_2\right)=0,
\end{array}
\eqno(14)
$$
and 
$$
f_2=\frac{f_{1x}+f_1^2-2c_1f_1-c_1^2-2c_2^2}{2(2c_2-f_1)},\qquad 
f_3=\frac{f_{1x}-f_1^2+2(2c_2-c_1)f_1-c_1^2-2c_2^2}{2(f_1-2c_2)}.
\eqno(15)
$$
The equation (14) can be written in more compact form ,
$$
\begin{array}{l}
2FF_{xx}-F_x^2+4\alpha FF_x-\left((F-3c_2+2\alpha)^2-3(\alpha^2+2c_2^2-2c_2\alpha)\right)\times\\
\left(3(F-c_2)^2-\alpha^2-6c_2^2+6\alpha c_2\right)=0,
\end{array}
\eqno(16)
$$
where $F=2c_2-f_1$, $\alpha=c_2-c_1$. 

Substituting (14-15) into the (8) one get 
$$
\begin{array}{l}
f_{1t}=-2c_1\left(f_1^2-2c_1f_1+2f_1f_2-4c_2f_2-c_1^2-2c_2^2\right)\\
f_{2t}=2c_2\left(f_2^2-2c_2f_2+2f_1f_2-4c_1f_1-c_2^2-2c_1^2\right).
\end{array}
$$
Thus, if $c_1=c_2=0$ then we get stationary solutions of the mBq equation. 

The equations for the $f_{1x}$ ($f_{2x}$) is compatible with the equation for the 
$f_{1t}$ ($f_{2t}$) if $c_1=c_2$ or if 
$$
F_t=2c_1F_x.
$$
Therefore, if $c_1\ne c_2$ then $F=F(\xi)$ with $\xi=x+2c_1t$ and $F(\xi)$ should be solution of 
the (16) with substitution $F_x\to F_{\xi}$.

Thus $H_{_{II }}$ (13) is parasuperhamiltonian if
$$
\begin{array}{l}
\frac{2}{3}a_1=f_{1x}-\left(f_1+c_1\right)^2,\qquad \frac{2}{3}a_2=-f_{1x}-\left(f_1+c_1\right)^2\\
\frac{2}{3}a_3=4c_1f_1-2f_1f_2-2f_2^2+2c_1^2,
\end{array}
$$
where $f_1$ and $f_2$ are defined by (14), (15).
\section{Complete PSUSY algebra.}

As we have seen, the usual PSUSY (5) is valid for the special kind of potentials only. 
On the other hand, the complete PSUSY algebra must be connected  with parasuperhamiltonian 
$H_{_I}$ (11) rather than $H_{_{II}}$ (13). This because $H_{_{II}}$ is connected with 
the auxiliary dynamical  problem whereas all information about mBq equation is contained 
in $H_{_I}$. Using this operator one can obtain the complete PSUSY algebra. In contrast 
to SUSY algebra (3) the complete PSUSY algebra is defined by 
superhamiltonian $H_{_I}$ and three parasupergenerators,
$$                              
Q_1=\left(\begin{array}{ccc}      
0&0&0\\                       
q_1&0&0\\                       
0&q_2&0                         
\end{array}\right),            
\qquad                             
Q_2=\left(\begin{array}{ccc}                                    
0&0&q_3\\                       
0&0&0\\                       
0&q_2&0                         
\end{array}\right),            
\qquad
Q_3=\left(\begin{array}{ccc} 
0&0&q_3\\                       
q_1&0&0\\                       
0&0&0                         
\end{array}\right),                                    
$$                              
whereas for the SUSY (3) it is enough to have superhamiltonian and two 
supergenerators $Q$ and $Q^+$. 

There are useful relations,
$$
\begin{array}{l}
M^3=H_{_I},\qquad M^2=Q_1^2+Q_2^2+Q_3^2,\qquad \{Q_i,Q_k\}=M^2,\qquad i\ne k\\
Q_1Q_2Q_3= Q_2Q_3Q_1= Q_3Q_1Q_2=Q_1^2=Q_2^2=Q_3^2=0\\
Q_1Q_3Q_2+Q_2Q_1Q_3+Q_3Q_2Q_1=2H_{_I},\qquad [Q_k,H_{_I}]=0,
\end{array}
\eqno(17)
$$
with $i$, $k=1,2,3$. 

(17) is para-generalization of the (3). To proof three last 
equations one need to use the intertwining relations,
$$
q_1L_1=L_2q_1,\qquad q_2L_2=L_3q_2,\qquad q_3L_3=L_1q_3.
$$ 

\section {Conclusion.}
Thus PSUSY algebra is underlie of algebraic structure of link between Bq and 
modified Bq equations. Now it is easy to find Miura transformation using the method from  the 
[1]. The results is well known (see [2]) so we omit them here. 

We conclude that parasupersymmetry is  useful not only in the quantum 
theory but in the theory of integrable systems. 
$$
{}
$$
{\bf Acknowledgement}
\newline
This work was supported by the Grant of Education Department of the 
Russian Federation,  No. E00-3.1-383.
$$
{}
$$
\centerline{\bf References}
\noindent
\begin{enumerate}
\item V.A. Andreev and M.V. Burova, Theor. Math. Phys., {\bf 85}, 376 (1990).
\item F. Gesztesy, D. Race and R. Weikard, J. London Math. Soc (2) {\bf 47}, 321 (1993).
\item E. Witten, Nucl. Phys. B {\bf 185}, 513 (1981).
\item V.A. Rubakov and V.P. Spiridonov, Mod. Phys. Lett. A {\bf 3}, 1337 (1988).
\item J. Beckers and N. Debergh, Nucl. Phys. B {\bf 340}, 767 (1990).
\item  A.A. Andrianov and M.V. Ioffe, Phys. Lett. B {\bf 255}, 543 (1990).
\item J. Beckers and N. Debergh, J. Phys. A {\bf 23}, L751 (1990).
\item V.B. Matveev and M.A. Salle, {\em Darboux Transformations and Solitons},
Springer, Berlin (1991).
\item  A.A. Andrianov, N.V. Borisov and M.V. Ioffe, Phys. Lett. A {\bf 105}, 19 (1984).
\item V.B. Matveev, J. Math. Phys. {\bf 35}, No. 6, 2965 (1994).
\end{enumerate}

\vfill
\eject

\end{document}